# Energy Dissipation and Asymmetric Excitation in Hybrid Waveguides for Routing and Coloring

*Xianguang Yang[1],\*, Long Wen[1], Jiahao Yan[1], Yanjun Bao[1], Qin Chen[1], Andrea Camposeo[2], Dario Pisignano[2,3] and Baojun Li[1],\**

[1] Institute of Nanophotonics, Jinan University, Guangzhou 511443, China

[2] NEST, Istituto Nanoscienze-CNR and Scuola Normale Superiore, Piazza S. Silvestro 12, I-56127 Pisa, Italy

[3] Dipartimento di Fisica, Università di Pisa, Largo B. Pontecorvo 3, I-56127 Pisa, Italy

\*E-mail: xianguang@jnu.edu.cn and baojunli@jnu.edu.cn

**Abstract:** The delivery of optical signals from an external light source to a nanoscale waveguide is highly important for the development of nanophotonic circuits. However, the efficient coupling of external light energy into nanophotonic components is difficult and still remains a challenge. Herein, we use an external silica nanofiber to light up an organic-inorganic hybrid nano-waveguide, namely a system composed of a polymer filament doped with $MoS_2$ quantum dots. Nanofiber-excited nano-waveguides in a crossed geometry are found to asymmetrically couple excitation signals along two opposite directions, with different energy dissipation resulting in different colors of the light emitted by $MoS_2$ quantum dots and collected from the waveguide terminals. Interestingly, rainbow-like light in the hybrid waveguide is achieved by three-in-one mixing of red, green, and blue components. This hetero-dimensional system of dots-in-waveguide represents a significant advance towards all-optical routing and full-color display in integrated nanophotonic devices.

Energy dissipation and energy transport are highly important in view of managing and optimizing energy-conversion processes at the micro- and nano-scale and of developing efficient photonic devices [1]. In particular, the microscopic mechanisms at the base of energy dissipation in either plasmonic [2] or dielectric [3] one-dimensional waveguides have been investigated for their relevance in the fields of optics and photonics. In one-dimensional waveguides, the main mechanisms contributing to the attenuation of the propagating light are self-absorption, light scattering by bulk and surface defects, and coupling to the substrate, the latter being increasingly relevant by decreasing the transverse size of the waveguide, whereas self-absorption can play a significant role in active waveguides, where light absorbing/emitting nanoparticles or molecules are embedded in the waveguides [4].

Two-dimensional materials offer new insight in this respect. Recently, molybdenum disulfide ($MoS_2$) has attracted an increasing interest due to its remarkable properties as semiconductor and for photonics [5]. The bandgap of $MoS_2$ could be well tailored by carefully controlling the number of layers stacked in its thickness. Few- and monolayer $MoS_2$ show a bandgap energy in the range 1.2-2.2 eV, tailorable by the number of layers and the lack of dangling bonds [6, 7]. Once the size of $MoS_2$ nanomaterial goes below 10 nm and becomes comparable to exciton Bohr radius, the bandgap of layered $MoS_2$ can be further engineered and varied because of quantum confinement effect. Thus, monolayer $MoS_2$ quantum dots (QDs) can show intriguing optical properties and are highly suitable for achieving nanophotonic components [8]. In addition, monolayer $MoS_2$ QDs have been proved to be more efficient in photoluminescence (PL) than their few-layer counterpart [9]. However, little attention has been paid so far to the possible combination of $MoS_2$ QDs with polymer systems suitable to build photonic devices [10, 11] and especially to the possible incorporation of active $MoS_2$ components into one-dimensional polymer waveguides.

Waveguides with sub-wavelength diameters, physically drawn from melts and/or solutions of polymer matrices, can be functionalized by QDs and used as building blocks of photonic networks, with the potentiality to be a scalable, novel platform for routing of light at the nanoscale [12]. This class of hybrid systems would be cavity-free, and it may show enhanced and broadband evanescent coupling of the optical signal from fluorescent QDs [13]. In this work, an organic polymer waveguide doped by monolayer $MoS_2$ QDs (hereinafter termed 'hybrid waveguide') is found to lead to combined subwavelength field localization and near-field broadband coupling without resonant conditions and low optical losses. Nonresonant interactions in these hybrid waveguides [14] make them highly suitable for room-temperature operation and are appealing for future energy technologies, such as energy-saving photonic circuits and light sources [15].

Indeed, the wave-guiding performance is highly important for hybrid waveguides to realize optical interconnections. The Stokes shift between absorption and emission of QDs in the hybrid waveguide largely affects these properties. Generally, the large Stokes shift (~100 meV) features low self-absorption for QD-emitted light, thus potentially leading to low optical losses. Low self-absorption is also useful for obtaining laser emission from hybrid waveguides exhibiting stimulated radiation [16, 17]. In addition, a large Stokes shift facilitates experimental studies of the excitation and emission properties of hybrid waveguides, allowing different signals to be easily separated and the microscopic mechanisms at the base of energy dissipation and transport to be better clarified. While a lot of efforts have been made to reduce self-absorption [18], exploiting such effect for adding new functionalities to light emitting nano-waveguides is an unexplored route.

We experimentally studied an exotic system that counterintuitively uses the self-absorption related energy dissipation in hybrid waveguides to feature a novel photonic function, namely, to generate color gradients in the emission. The spatially varying full-color emission along the hybrid waveguide can be considered as mimicking a rainbow at nanoscale. Up to now, multicolor emissions were achieved with multicomponent materials, where multiple luminescent materials lead to the respective emission colors [19]. In such systems, it is difficult to achieve a micro- or nanoscale spatial variation of the color of the emission because the luminescent materials are typically uniformly distributed in the nano-waveguides. Moreover, to reduce the absorption of high energy photons by the narrow bandgap material, materials with different luminescence properties need to be spatially separated [20], thus increasing the fabrication complexity. In our waveguides, instead, there is a single emitting nanomaterial, while the propagation direction, the wavelength and the spatial variation of the intensity of the excitation light are synergistically exploited to vary continuously the local color of the out-coupled light, overall achieving a compact and robust, nanostructured photonic system [21, 22].

Figure 1a presents the flame assisted drawing for fabricating nanofibers from a silica tapered fiber with two oppositely exerting forces. The nanofiber is positioned into the outer flame, where oxygen is sufficient for combustion. The nozzle diameter of ethanol flame is about 6 mm. Figure 1b shows an optical micrograph of an as-fabricated nanofiber from a 20-μm-diameter silica tapered fiber (see also the Supporting Information). This type of nanofiber has been proved to offer high efficiency and broadband coupling for near-field applications [23]. Figure 1c illustrates the preparation of a solution containing polyvinylpyrrolidone (PVP) and $MoS_2$ QDs (Figure S1 and S2), for drawing a hybrid nano-waveguide. Figure 1d shows the physical drawing process, leading to the nano-waveguide from PVP-QDs solution by using a silica fiber (see Supporting Information). This drawing method is highly cheap to implement, and it does not require high-velocity jets or electric fields as in electrospinning. Furthermore, it allows the hybrid nano-waveguide to be *in-situ* controlled and robustly manipulated at single-filament level.

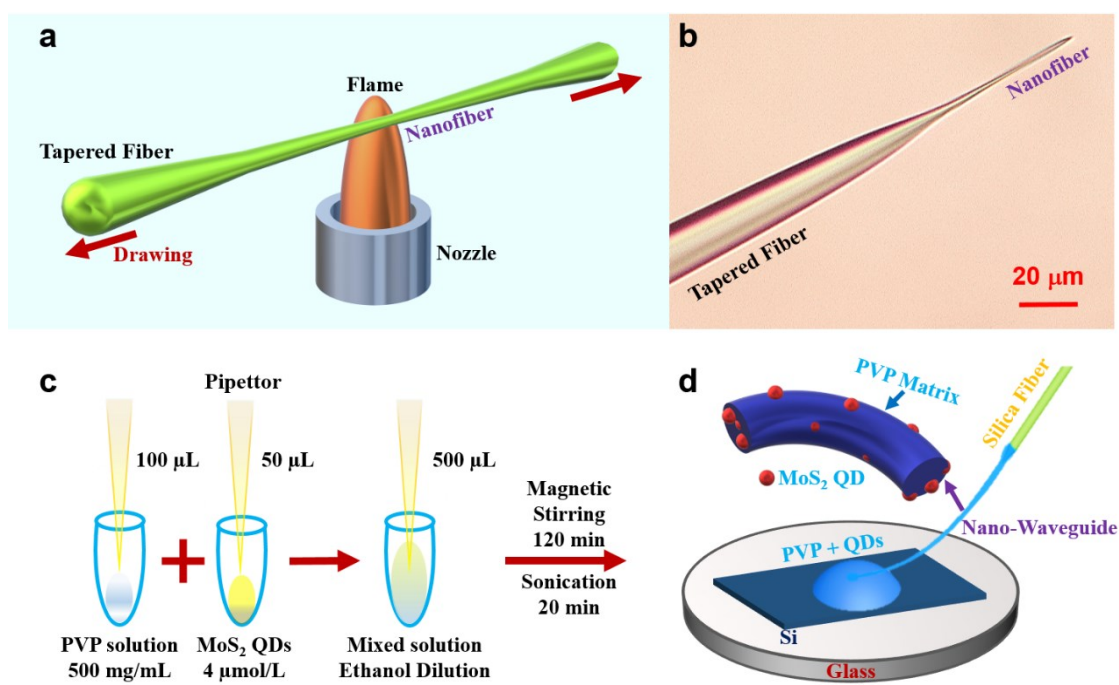

**Figure 1. Fabrication of nanofibers and of nano-waveguides.** (a) Schematic diagram of the flame drawing for producing nanofibers out from a silica tapered fiber. The nozzle diameter of ethanol flame is about 6 mm. (b) Optical micrograph of an as-fabricated nanofiber, generated from a 20 μm diameter silica tapered fiber. (c) Scheme of the solution preparation (PVP and MoS$_2$ QDs in ethanol). Then, magnetic stirring for 120 min and sonication for 20 min are performed to form a uniform solution with an appropriate viscosity for drawing nano-waveguide. (d) Direct drawing of the nano-waveguide from the prepared solution. Inset: scheme of the fabricated nano-waveguide with MoS$_2$ QDs embedded in the PVP matrix.

To investigate the morphology of as-fabricated nanofiber, Scanning Electron Microscope (SEM) measurements were performed. The obtained SEM image of typical silica nanofiber is shown in Figure 2a. The smooth and uniform tapering of the silica nanofiber is highly suitable to favor high efficiency (~40%) of the coupling of light towards the hybrid nano-waveguide [23]. Figure 2b shows the SEM image of a hybrid nano-waveguide with average diameter at 300 ± 10 nm. The residues on the surface of the nano-waveguide are attributed to gold sputtering for SEM measurements. For assessing further the composition and structure of the hybrid nano-waveguides, Transmission

Electron Microscopy (TEM) equipped with an energy-dispersive X-ray (EDX) spectrometer was used. Figure 2c shows the TEM image of the 300 nm diameter nano-waveguide with smooth surface. The $MoS_2$ QDs were successfully incorporated, as highlighted by yellow arrows, and no aggregation was observed (see the Supporting Information and Figure S3). The diameter variation is of 4 nm along a waveguide length of 2 µm, as assessed by atomic force microscopy (Figure S4). Overall, the relatively uniform diameter and the smooth surface are beneficial to transport light energy [24], and for broadband light confinement because more than 60% of the power of the fundamental mode is confined in the core of the nano-waveguide for wavelength < 650 nm (see Supporting Information and Figure S5 for details).

The EDX spectrum shown in Figure 2d confirmed the presence of S (11.85 Atomic %) and Mo (8.34 Atomic %) in the hybrid nano-waveguide, giving the stoichiometry ratio of 1.421. The difference of stoichiometry may be attributed to vacancies associated with scaling down the physical dimensions [5, 8, 25]. The above two elements come from the doped $MoS_2$ QDs. The other elements (Cu, C and O) come from the copper micro-grid with carbon membrane used during TEM measurements. The doped concentration of $MoS_2$ QDs in the hybrid nano-waveguide is estimated to be of $3.6 \times 10^3$ $\mu m^{-3}$ (see also Supporting Information). To study the QD luminescence, dark-field fluorescent measurements were carried out on the hybrid nano-waveguide. Figure 2e shows the fluorescence micrograph of a hybrid nano-waveguide with the illumination of a laser. The apparent diameter of ~1 µm (limited by the spatial resolution of fluorescence microscopy) is due to the signal from the incorporated $MoS_2$ QDs, which were successfully excited. Notably, the incorporation of $MoS_2$ QDs in the PVP matrix can produce synergistic effects. On one hand the QDs endow the PVP matrix novel fluorescence, on the other hand the PVP matrix can protect the QDs from corrosion/oxidation [26].

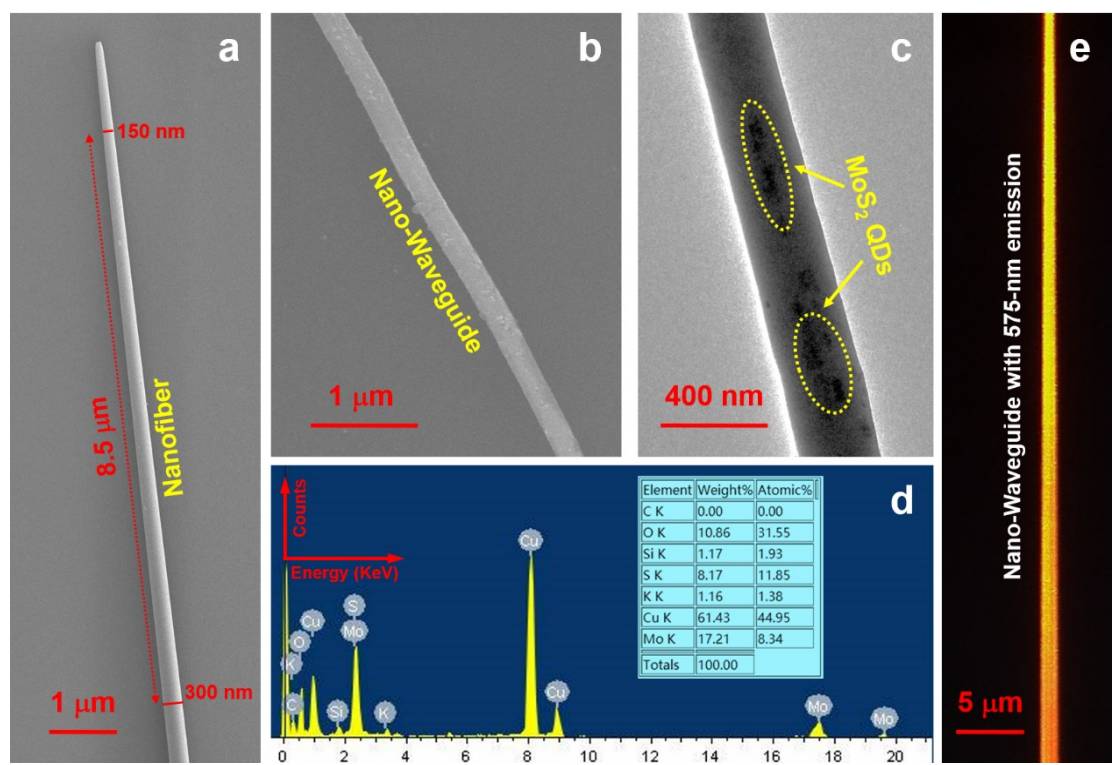

**Figure 2. Microscopic characterization of nanofibers and of nano-waveguides.** (a) SEM image of a silica nanofiber. (b) SEM image of a 300 nm diameter nano-waveguide. (c) TEM image of a nano-waveguide. Yellow arrows show the embedded $MoS_2$ QDs. (d) EDX spectrum of the nano-waveguide in (c), the inset shows the element percentage. (e) Fluorescence micrograph of a nano-waveguide with emission wavelength at 575 nm.

PL is highly desirable and important for all-optical routing and wavelength-conversion applications. The PL of a hybrid nano-waveguide was studied by using a silica nanofiber to guide the excitation light, which was launched from an external semiconductor laser diode. Figure 3a shows the bright-field optical micrograph of a studied structure, comprising a silica nanofiber crossed with a hybrid nano-waveguide. The inset shows the good connection, which is important for effective optical coupling. The average diameter of the nanofiber and of the nano-waveguide is about 200 and 300 nm, respectively, as derived by SEM measurements. To suppress light energy leakages into the substrate supporting the nano-waveguide, a low refractive index layer made of magnesium fluoride was deposited on the silicon wafer underneath [27]. Figure 3b shows the dark-field micrograph of the structure excited with a 532-nm 100-nW green laser guided by nanofiber. The white arrow indicates the propagation direction. Upon guiding the 532-nm laser light to the cross junction (position M), the doped QDs of nano-waveguide were excited due to evanescent coupling [28]. The coupled green laser is unequally transmitted to the two distal ends (D1 and D2) of the nano-waveguide, as confirmed by the green light spots shown in insets D1 and D2 and Figure S6. Most of the coupled green laser was guided to the nano-waveguide end D1, hence the QDs in segment M-to-D1 were excited more efficiently than in the segment M-to-D2. This asymmetric light transmission can be attributed to symmetry breaking by a 55-degree cross angle between the two fiber systems, resulting in directional light coupling from the tapered nanofiber to the nano-waveguide branch that forms a smaller angle with the direction of propagation of the incident light, as demonstrated by finite-difference time-domain (FDTD) simulations (see Figure S6) and experiments [28-30]. Such directional coupling may find potential applications for asymmetric routing of light [31].

To obtain the PL intensity variation and dependence on the propagation distance, the microscopically-resolved emission (micro-PL) was collected along the nano-waveguide. Figure 3c and 3d show the measured PL for distances along M-to-D1 (c) and along M-to-D2 (d), respectively. The PL intensity is gradually decreasing upon increasing the distance, which is due to the optical loss associated with energy dissipation. To characterize the optical loss, the excitation position M of the tapered nanofiber was moved along the nano-waveguide with 5 μm steps by micro-manipulation, while the integral intensities at the cross junction ($I_{body}$, position M) and distal end ($I_{end}$, position D) were measured for different distances between the nano-waveguide tip and the tapered nanofiber. Figure 3e shows the intensity ratio ($I_{end}/I_{body}$) as a function of the guiding distance ($d$) with a logarithmic scale. An optical loss factor of $\alpha = 0.2107$ dB/μm was calculated by fitting as equation: Log ($I_{end}/I_{body}$) $= -\alpha d$. The energy dissipation mainly comes from the absorption of the light from the excitation laser, fluorescence self-absorption by un-excited QDs, and Rayleigh scattering from the embedded nanoparticles (see Supporting Information for additional discussion about the propagation losses). The more self-absorption, the larger energy dissipation, resulting in stronger spectral red shifts. More specifically, fluorescence light collected at the end of segment M-to-D2 mainly comes from QDs located nearby the crossing point M (Fig. 3b), due to the rapidly decaying spatial behavior of the excitation light coupled from M into the nano-waveguide. For this reason, such fluorescence component undergoes self-absorption along the entire length of the M-to-D2 segment. Instead, in segment M-to-D1 the more intense excitation light is able to reach also QDs far from the crossing point M and closer to the end of the segment (Fig. 3b). The fluorescence light so-emitted from areas closer to the segment end D1 are significantly less affected by self-absorption losses. Overall, the emission collected from the distal end D2 is consequently red-shifted with respect to that collected from the distal end D1, leading to different colors of light emission at the two waveguide terminals. Figure 3f shows the relationship between the emission peak wavelength and the guiding distance. The original peak wavelength of 575 nm red shifts to 580 nm as the distance increases up to 30 μm. It can be anticipated that a new degree of freedom for tailoring the emission wavelength is created by controlling the waveguide length. In addition, Fabry-Pérot micro-cavity lasing could occur, if the nano-waveguide terminals have a neat and clean cross section [32] and sufficient optical gain is achieved by stimulated emission in order to overcome the cavity losses.

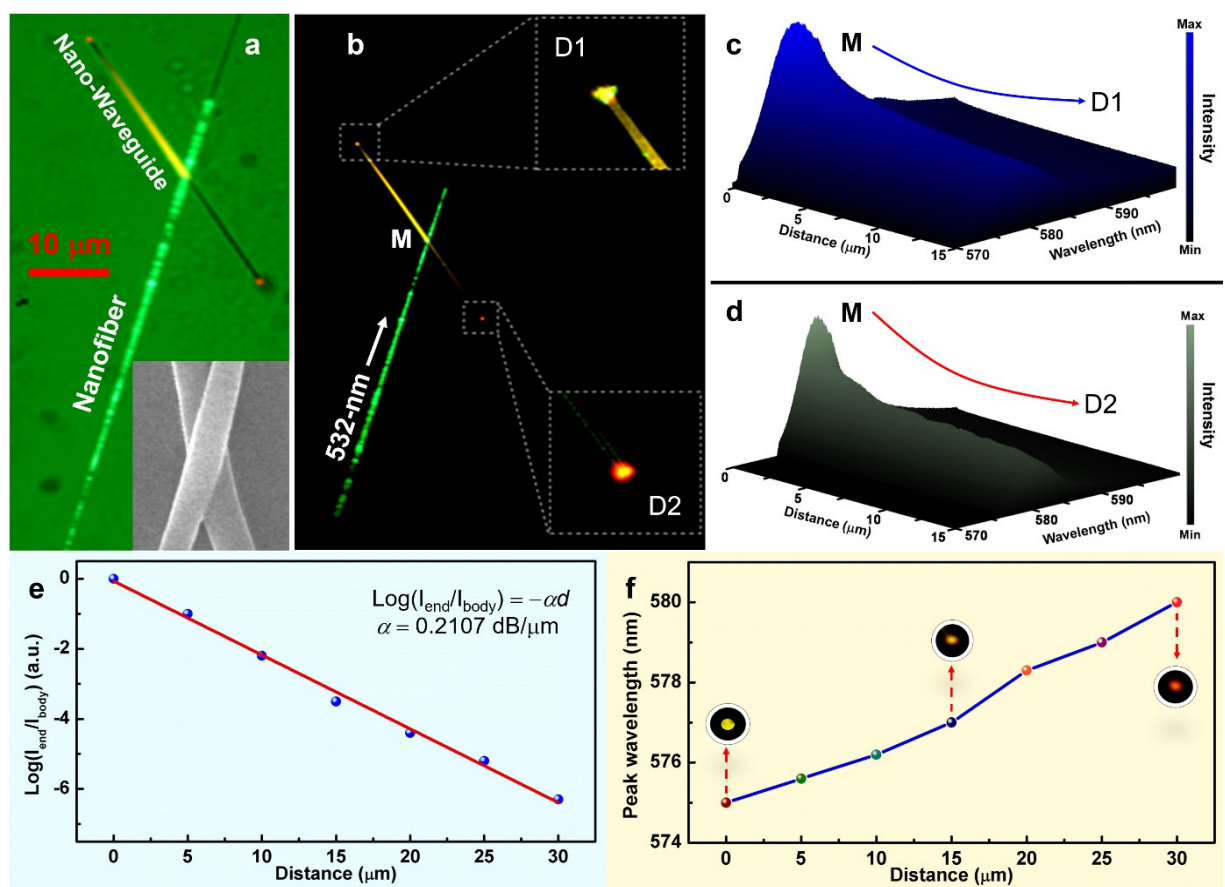

**Figure 3. Nanofiber-asymmetrically excited nano-waveguide.** (a) Bright-field micrograph of a nanofiber crossed nano-waveguide structure. Inset: connection between nano-waveguide and nanofiber. b) Dark-field micrograph of the structure excited with 532-nm green laser guided by nanofiber. The insets present the two distal ends (D1 and D2) of the nano-waveguide. (c, d) Micro-PL evolution along the distance for (c) middle M to distal D1 and (d) M to D2, respectively. (e) Logarithmic plot of intensity ratio ($I_{end}/I_{body}$) as a function of the guiding distance (*d*) between the cross junction (excited spot) and the distal end (emission output). (f) The peak wavelength at different guiding distance. Insets are optical micrograph.

Figure 4a shows a bright-field optical micrograph of a studied structure, with two tapered silica nanofibers tightly contacting with the same nano-waveguide due to van der Waals interaction. The inset shows a SEM image of the studied nano-waveguide. The tapered fibers (1 and 2) are used to guide 532 and 473 nm laser light, respectively. Figure 4b shows a dark-field optical micrograph of the structure in (a), which was simultaneously excited by the green and the blue laser light with 50-nW-power. White arrows indicate the propagation directions of the excitation laser light. When the excitation lights evanescently coupled into the nano-waveguide, the $MoS_2$ QDs were excited and emit fluorescence. The excitation and emission lights undergo a wavelength-converted wave-guiding process [33], exhibiting colorful emission along the nano-waveguide. The inset of Fig. 4b shows a magnified view highlighting the resulting, iridescent colors, eventually leading to white light generation (see Figure S7). The colorful emission can be attributed to three-in-one mixing of red (R ~575 nm), green (G ~532 nm), and blue (B ~473 nm) light in the hybrid nano-waveguide. This strategy demonstrated an additive behavior of RGB primary colors at nanometer scale, mimicking 1D nanoscale *rainbow*. Compared with structural color generation, the additive color of nanoscale RGB lights do not request any complex design of structured interfaces [34]. Furthermore, the low energy consumption (50 nW level) could make these systems applicable for energy-saving, full-color displays. The spatial distribution and time stability of light emission in nano-waveguides are also important for practical applications. In order to evaluate the spatial distribution of PL, Figure 4c shows confocal PL spectra collected from an area containing a 12-μm-long nano-waveguide, under 532 nm excitation. The region has a size of $2.5 \times 12$ μm$^2$, imaged with spatial resolution of 0.5 μm. The inset of Fig. 4c shows the spatial distribution of the PL intensity at a peak wavelength of 575 nm. A uniform distribution was observed from confocal PL mapping, where red dotted lines indicate the outline of the luminescent nano-waveguide. For the photo-stability, Figure 4d shows time dependent PL intensity at 575 nm, where red data were measured from individual positions on a nano-waveguide (the green spot indicates the sampling point). The PL intensity decreased to 50% of initial value during the first 30 min, which is due to the associated photo-chemistry. After 60 min, about 37.5% of initial intensity still remained, which is benefited by the PVP matrix protecting the doping QDs from oxidation. This time-related photo-stability well complements the characterization of absorption-stability for $MoS_2$ QDs [35].

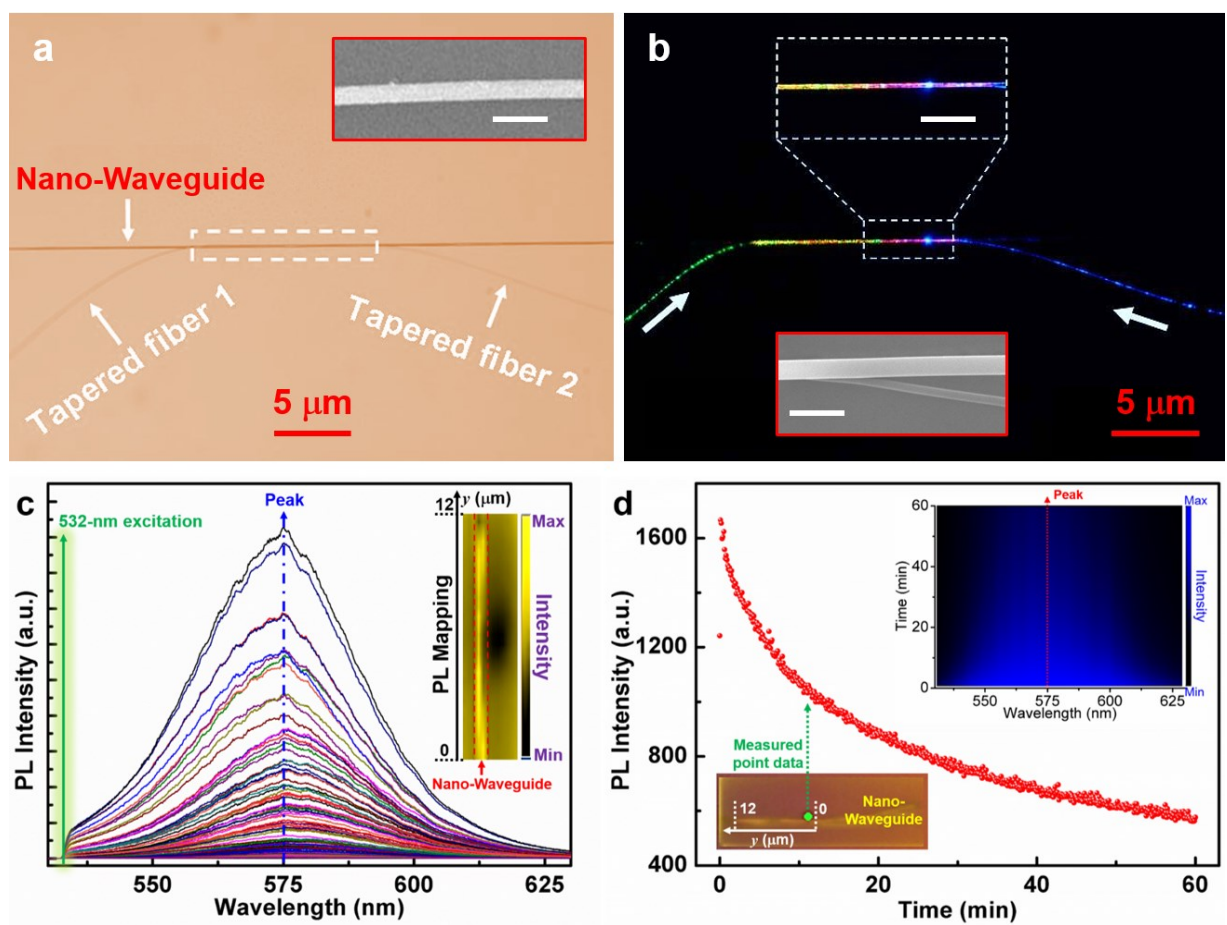

**Figure 4. Colorful emission, spatial and temporal PL mapping.** (a) Bright-field optical micrograph of two tapered nanofibers coupling with the same nano-waveguide. Inset: SEM image of 300 nm diameter nano-waveguide. (b) Dark-field optical micrograph of the structure in (a) simultaneously excited by both green and blue lasers. Insets: magnified view of the colorful emission in the nano-waveguide. SEM image shows the good connection between nano-waveguide and tapered fiber 2. (c) Confocal PL spectra collected from an area (2.5 × 12 µm$^2$, spatial resolution of 0.5 µm), containing a 12-µm-long nano-waveguide. The inset shows the spatial distribution of PL intensity at 575 nm. (d) Time dependent PL intensity at 575 nm measured from an individual point (green spot). Inset: time-correlated mapping of the PL spectrum (interval of 6 s).

To summarize, organic-inorganic hybrid nano-waveguides were physically-drawn from an organic PVP matrix doped with MoS$_2$ QDs. Asymmetric routing of light energy was obtained by a 55-degree cross angle to break the symmetry of evanescent field coupling. Two different colors of light emissions at the nano-waveguide terminals were achieved because of self-absorption related energy dissipation. Three-in-one mixing of nanoscale RGB signals in nano-waveguides produced colorful emission as 1D rainbow. The uniform and stable emission may favor the use of hybrid nano-waveguides in various fields including optical routing in photonic circuits and devices [36], colorful displays and output coupling of laser emissions [37].

**Acknowledgements**

This work was supported by the National Natural Science Foundation of China (Nos. 11804120, 61827822, 11874029, 11774383 and 62005096), the Guangdong Basic and Applied Basic Research Foundation (2019A1515110414), and the Research Projects from Guangzhou (201804010468). D.P. acknowledges the Italian Minister of University and Research PRIN 2017PHRM8X project ("3D-Phys"), and the PRA_2018_34 ("ANISE") project from the University of Pisa.

**Supporting Information**

Materials and methods; fabrications of nanofiber and nano-waveguide; characterizations and additional discussion about the propagation losses; numerical simulation; white light generation.

**References**

(1) Pop, E. Energy Dissipation and Transport in Nanoscale Devices. *Nano Res.* **2010**, *3*, 147-169.

(2) Guo, X.; Ma, Y.; Wang, Y.; Tong, L. Nanowire Plasmonic Waveguides, Circuits and Devices. *Laser Photonics Rev.* **2013**, *7*, 855-881.

(3) O'Carroll, D.; Lieberwirth, I.; Redmond, G. Melt-Processed Polyfluorene Nanowires as Active Waveguides. *Small* **2007**, *3*, 1178-1183.

(4) Liu, H.; Edel, J. B.; Bellan, L. M.; Craighead, H. G. Electrospun Polymer Nanofibers as Subwavelength Optical Waveguides Incorporating Quantum Dots. *Small* **2006**, *2*, 495-499.

(5) Huang, X.; Zeng, Z.; Zhang, H. Metal Dichalcogenide Nanosheets: Preparation, Properties and Applications. *Chem. Soc. Rev.* **2013**, *42*, 1934-1946.

(6) Mak, K. F.; Lee, C.; Hone, J.; Shan, J.; Heinz, T. F. Atomically Thin $MoS_2$: A New Direct-Gap Semiconductor. *Phys. Rev. Lett.* **2010**, *105*, 136805.

(7) Yang, X.; Li, B. Monolayer $MoS_2$ for Nanoscale Photonics. *Nanophotonics* **2020**, *9*, 1557-1577.

(8) Xu, Y.; Wang, X.; Zhang, W. L.; Lv, F.; Guo, S. Recent Progress in Two-Dimensional Inorganic Quantum Dots. *Chem. Soc. Rev.* **2018**, *47*, 586-625.

(9) Gopalakrishnan, D.; Damien, D.; Shaijumon, M. M. $MoS_2$ Quantum Dot-Interspersed Exfoliated $MoS_2$ Nanosheets. *ACS Nano* **2014**, *8*, 5297-5303.

(10) Wang, X.; Xing, W.; Feng, X.; Song, L.; Hu, Y. $MoS_2$/Polymer Nanocomposites: Preparation, Properties, and Applications. *Polym. Rev.* **2017**, *57*, 440-466.

(11) Portone, A.; Romano, L.; Fasano, V.; Di Corato, R.; Camposeo, A.; Fabbri, F.; Cardarelli, F.; Pisignano D.; Persano, L. Low-Defectiveness Exfoliation of $MoS_2$ Nanoparticles and Their Embedment in Hybrid Light-Emitting Polymer Nanofibers. *Nanoscale* **2018**, *10*, 21748-21754.

(12) Wang, P.; Wang, Y.; Tong, L. Functionalized Polymer Nanofibers: A Versatile Platform for Manipulating Light at the Nanoscale. *Light: Sci. Appl.* **2013**, *2*, e102.

(13) Yalla, R.; Le Kien, F.; Morinaga, M.; Hakuta, K. Efficient Channeling of Fluorescence Photons from Single Quantum Dots into Guided Modes of Optical Nanofiber. *Phys. Rev. Lett.* **2012**,

*109*, 063602.

(14) Gaio, M.; Saxena, D.; Bertolotti, J.; Pisignano, D.; Camposeo, A.; Sapienza, R. A Nanophotonic Laser on a Graph. *Nat. Commun.* **2019**, *10*, 226.

(15) Demir, H. V.; Nizamoglu, S.; Erdem, T.; Mutlugun, E.; Gaponik, N.; Eychmüller, A. Quantum Dot Integrated LEDs Using Photonic and Excitonic Color Conversion. *Nano Today* **2011**, *6*, 632-647.

(16) Yang, X.; Li, B. Laser Emission from Ring Resonators Formed by a Quantum-Dot-Doped Single Polymer Nanowire. *ACS Macro Lett.* **2014**, *3*, 1266-1270.

(17) Le Feber, B.; Prins, F.; De Leo, E.; Rabouw, F. T.; Norris, D. J. Colloidal-Quantum-Dot Ring Lasers with Active Color Control. *Nano Lett.* **2018**, *18*, 1028-1034.

(18) Ning, C.-Z.; Dou, L.; Yang, P. Bandgap Engineering in Semiconductor Alloy Nanomaterials with Widely Tunable Compositions. *Nat. Rev. Mater.* **2017**, *2*, 17070.

(19) Cerdán, L.; Enciso, E.; Martín, V.; Bañuelos, J.; López-Arbeloa, I.; Costela, A.; García-Moreno, I. FRET-Assisted Laser Emission in Colloidal Suspensions of Dye-Doped Latex Nanoparticles. *Nat. Photonics* **2012**, *6*, 621-626.

(20) Zhao, J.; Yan, Y.; Gao, Z.; Du, Y.; Dong, H.; Yao, J.; Zhao, Y. S. Full-Color Laser Displays Based on Organic Printed Microlaser Arrays. *Nat. Commun.* **2019**, *10*, 870.

(21) Yang, Z.; Albrow-Owen, T.; Cui, H.; Alexander-Webber, J.; Gu, F.; Wang, X.; Wu, T.-C.; Zhuge, M.; Williams, C.; Wang, P.; et al. Single-Nanowire Spectrometers. *Science* **2019**, *365*, 1017-1020.

(22) Liao, F.; Yu, J.; Gu, Z.; Yang, Z.; Hasan, T.; Linghu, S.; Peng, J.; Fang, W.; Zhuang, S.; Gu, M. Enhancing Monolayer Photoluminescence on Optical Micro/Nanofibers for Low-Threshold Lasing. *Sci. Adv.* **2019**, *5*, eaax7398.

(23) Kim, S.; Yu, N.; Ma, X.; Zhu, Y.; Liu, Q.; Liu, M.; Yan, R. High External-Efficiency Nanofocusing for Lens-Free Near-Field Optical Nanoscopy. *Nat. Photonics* **2019**, *13*, 636-643.

(24) Yang, X.; Li, Y.; Lou, Z.; Chen, Q.; Li, B. Optical Energy Transfer from Photonic Nanowire to

Plasmonic Nanowire. *ACS Appl. Energy Mater.* **2018**, *1*, 278-283.

(25) Chiu, N.-F.; Tai, M.-J.; Nurrohman, D. T.; Lin, T.-L.; Wang, Y.-H.; Chen, C.-Y. Immunoassay-Amplified Responses Using a Functionalized $MoS_2$-Based SPR Biosensor to Detect PAPP-A2 in Maternal Serum Samples to Screen for Fetal Down's Syndrome. *Int. J. Nanomed.* **2021**, *16*, 2715.

(26) Zhang, C.-L.; Yu, S.-H. Nanoparticles Meet Electrospinning: Recent Advances and Future Prospects. *Chem. Soc. Rev.* **2014**, *43*, 4423-4448.

(27) Zhang, R.; Yu, H.; Li, B. Active Nanowaveguides in Polymer Doped with CdSe-ZnS Core-Shell Quantum Dots. *Nanoscale* **2012**, *4*, 5856-5859.

(28) He, W.; Li, B.; Pun, E. Y.-B. Wavelength, Cross-Angle, and Core-Diameter Dependence of Coupling Efficiency in Nanowire Evanescent Wave Coupling. *Opt. Lett.* **2009**, *34*, 1597-1599.

(29) Li, W.; Gao, Y.; Tong, L. Crosstalk in Two Intersecting Optical Microfibers. *IEEE Photonics Technol. Lett.* **2019**, *31*, 1514-1517.

(30) Li, Y. J.; Yan, Y.; Zhang, C.; Zhao, Y. S.; Yao, J. Embedded Branch-Like Organic/Metal Nanowire Heterostructures: Liquid-Phase Synthesis, Efficient Photon-Plasmon Coupling, and Optical Signal Manipulation. *Adv. Mater.* **2013**, *25*, 2784-2788.

(31) Li, Y. J.; Yan, Y.; Zhao, Y. S.; Yao, J. Construction of Nanowire Heterojunctions: Photonic Function-Oriented Nanoarchitectonics. *Adv. Mater.* **2016**, *28*, 1319-1326.

(32) Kuehne, A. J.; Gather, M. C. Organic Lasers: Recent Developments on Materials, Device Geometries, and Fabrication Techniques. *Chem. Rev.* **2016**, *116*, 12823-12864.

(33) Yang, X.; Bao, D.; Li, B. Light Transfer from Quantum-Dot-Doped Polymer Nanowires to Silver Nanowires. *RSC Adv.* **2015**, *5*, 60770-60774.

(34) Goodling, A. E.; Nagelberg, S.; Kaehr, B.; Meredith, C. H.; Cheon, S. I.; Saunders, A. P.; Kolle, M.; Zarzar, L. D. Colouration by Total Internal Reflection and Interference at Microscale Concave Interfaces. *Nature* **2019**, *566*, 523-527.

(35) Dong, H.; Tang, S.; Hao, Y.; Yu, H.; Dai, W.; Zhao, G.; Cao, Y.; Lu, H.; Zhang, X.; Ju, H.

Fluorescent MoS2 Quantum Dots: Ultrasonic Preparation, Up-Conversion and Down-Conversion Bioimaging, and Photodynamic Therapy. *ACS Appl. Mater. Interfaces* **2016**, *8*, 3107-3114.

(36) Zheng, J. Y.; Yan, Y.; Wang, X.; Zhao, Y. S.; Huang, J.; Yao, J. Wire-on-Wire Growth of Fluorescent Organic Heterojunctions. *J. Am. Chem. Soc.* **2012**, *134*, 2880-2883.

(37) Li, Y. J.; Lv, Y.; Zou, C.-L.; Zhang, W.; Yao, J.; Zhao, Y. S. Output Coupling of Perovskite Lasers from Embedded Nanoscale Plasmonic Waveguides. *J. Am. Chem. Soc.* **2016**, *138*, 2122-2125.

## Supporting Information

## Energy Dissipation and Asymmetric Excitation in Hybrid Waveguides for Routing and Coloring

*Xianguang Yang[1],\*, Long Wen[1], Jiahao Yan[1], Yanjun Bao[1], Qin Chen[1], Andrea Camposeo[2], Dario Pisignano[2,3] and Baojun Li[1],\**

[1] Institute of Nanophotonics, Jinan University, Guangzhou 511443, China

[2] NEST, Istituto Nanoscienze-CNR and Scuola Normale Superiore, Piazza S. Silvestro 12, I-56127 Pisa, Italy

[3] Dipartimento di Fisica, Università di Pisa, Largo B. Pontecorvo 3, I-56127 Pisa, Italy

\*E-mail: xianguang@jnu.edu.cn and baojunli@jnu.edu.cn

**Materials and methods; fabrications of nanofiber and nano-waveguide; characterizations and additional discussion about the propagation losses; numerical simulation; white light generation.**

**Materials:**

Polyvinylpyrrolidone (PVP, 98%) and ammonium tetrathiomolybdate [$(NH_4)_2MoS_4$, 99%] were commercially available from Alfa-Aesar (Shanghai, China) and J&K Chemical Ltd. (Beijing, China), respectively. Glycerol (99%) available from Sigma-Aldrich (Shanghai, China) can control the viscosity of the PVP ethanol solution, which prevents shrinking of the hybrid nano-waveguides upon solvent evaporation. The materials were used as received without further purification.

**Monolayer $MoS_2$ quantum dots (QDs):**

QDs were synthesized from the precursor of $(NH_4)_2MoS_4$ by a hydrothermal method. First, an amount of 10 mg of $(NH_4)_2MoS_4$ was added into a 40 mL of PVP solution (1 mg/mL) for dissolution via ultra-sonication. Once dissolved completely, an amount of 400 μL of $N_2H_4$ (80%) acting as the reductive agent was mixed with the above solution. Then, the obtained mixture was transferred into a 50 mL of Teflon-lined autoclave and kept at 120 °C for 5 h. When the solution was cooled down to room temperature, the impurities were filtered by a 200 nm micro-porous membrane for filtration of the unreactive residue. Finally, the solution was concentrated and further purified to remove the residual salts by ultra-filtration (Molecular Weight, 100 k). The concentration of synthesized $MoS_2$ QDs is 4 µmol/L, the average particle size is 5 nm (standard deviation of the size distribution 0.3 nm, Figure S1) with a thickness of 0.7 nm (Figure S2) that confirmed the formation of monolayer QDs. The emission wavelength is peaked at 575 nm.

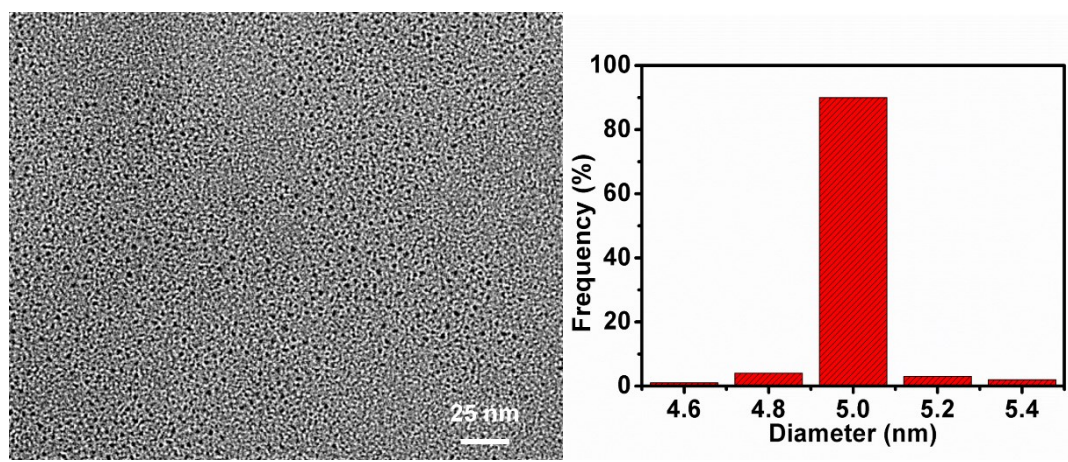

**Figure S1.** TEM image and diameter distribution of $MoS_2$ QDs.

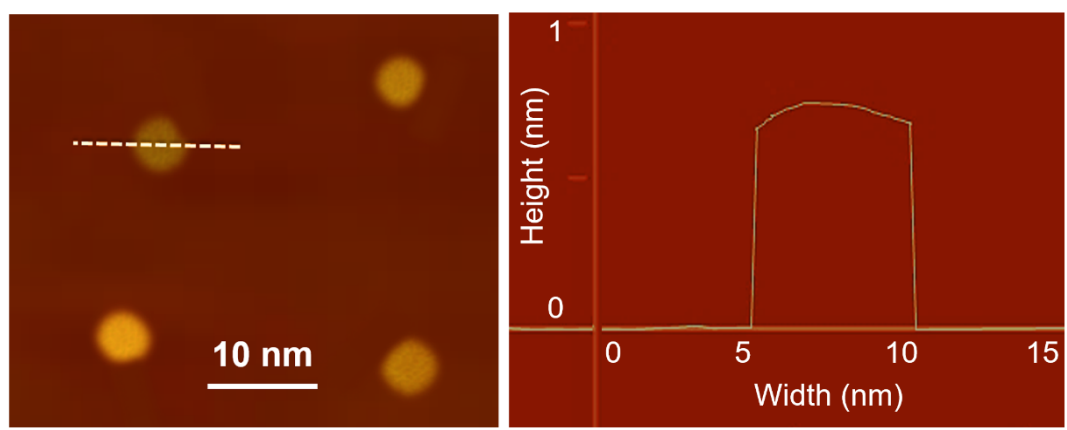

**Figure S2.** Atomic Force Microscope image (left image) and the height profile (right plot) of single QD.

**Fabrication of nanofiber:**

A substrate of silicon wafer was cleaned in oxygen plasma for 8 min. A layer (100 nm thickness) of magnesium fluoride ($MgF_2$) was coated on the silicon wafer to suppress optical leakages into the substrate. The tapered fiber tailed with nanofiber was fabricated by a flame-heating and mechanically-pulling method from a single mode standard telecommunication fiber (Corning Inc., USA). The core and cladding diameters of the Corning fiber are 10 and 125 µm, respectively, with connector type FC/PC. The jacket at the fiber end was cleaved and sonicated in acetone and DI water was used to remove residues form the surface around the cladding layer. The naked fiber was heated for about 35 s until reaching its melting point. The outer flame temperature was about 480 °C. The melting fiber was horizontally pulled at a rate of ~4 mm/s, the fiber diameter was gradually reduced from 125 to 2 µm, and the tapered length was about 3 mm. Finally, the stretching rate was increased up to ~15 mm/s until the melting fiber broken with a nanofiber. The resulted nanofiber was then thoroughly washed with DI water.

**Fabrication of nano-waveguide:**

Organic-inorganic hybrid nano-waveguides were fabricated as follows. First, PVP (3 g) was dissolved in anhydrous ethanol (5 mL) and glycerol (1 mL) to form a homogeneous solution. $MoS_2$ QDs (50 µL) suspended in ethanol with concentration of 4 µmol/L were added to the PVP solution (100 µL), and then diluted by 500 µL of ethanol. The mass of PVP ($M_{PVP}$) in 100 µL PVP-solution can be calculated by $M_{PVP}$ = 3 g / (5+1) mL × 100 µL. The volume of PVP ($V_{PVP}$) can be calculated by $V_{PVP} = M_{PVP} / D_{PVP}$, where $D_{PVP}$ is the density of PVP ($D_{PVP}$ = 1.144 g/cm³). The number of QDs ($N_{QD}$) can be calculated by $N_{QD}$ = 50 µL × 4 µmol/L × $N_A$, where $N_A$ is Avogadro constant ($N_A$ = 6.02×10²³). Therefore, the doped concentration of $MoS_2$ QDs ($C_{QD}$ ~3.6×10³ µm⁻³) can be calculated by $C_{QD} = N_{QD} / V_{PVP} = N_{QD} / (M_{PVP} / D_{PVP})$.

The mixture solution was magnetic stirred at room temperature for 120 min and then ultra-sonicated for 20 min to yield a uniform solution with proper viscosity for drawing nano-waveguides. The tip of a silica fiber (tip diameter, ~20 μm) was immersed into the solution for 5–10 s and then pulled out with a speed of 1–3 m/s, resulting a $MoS_2$ QDs embedded PVP wire between the solution and fiber tip due to the rapid evaporation of ethanol. After drying in air for about 2 h, solid-state nano-waveguides containing organic PVP and inorganic QDs were obtained. The diameter of the hybrid nano-waveguides are ranging from 200 to 400 nm. The doping concentration of $MoS_2$ QDs in the hybrid nano-waveguides can be tailored ranging from $2\times10^3$ $\mu m^{-3}$ to $4\times10^3$ $\mu m^{-3}$ without any appreciable aggregation, by controlling the content of added QDs during the fabrication procedure. A systematic study on 100 nano-waveguides shows that 95 nano-waveguides do not feature any appreciable aggregation. The yield is so estimated to be of 95%. A representative TEM image of nano-waveguide with QDs aggregation is presented as Figure S3. Furthermore, the fabrication setup is nozzleless and inexpensive to implement without requiring high-velocity jet or electric field. Robust waveguiding performance was obtained with QD-concentration of $2 \times 10^3$ $\mu m^{-3}$ to $4 \times 10^3$ $\mu m^{-3}$. For higher QD-concentration ($> 4 \times 10^3$ $\mu m^{-3}$), the waveguiding performance will be decreased due to high self-absorption of the emission light, as well as light scattering. For low QD-concentration ($< 2 \times 10^3$ $\mu m^{-3}$), the waveguiding performance will also be decreased due to low photoluminescence intensity, resulting from less emitters.

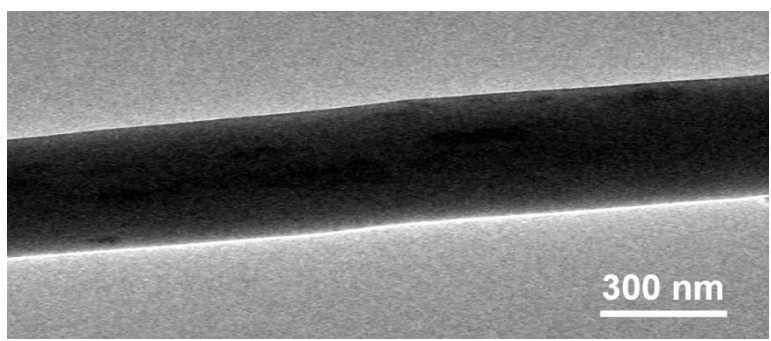

**Figure S3**. TEM image of nano-waveguide with QDs aggregation.

**Microscopic Characterization:**

Micro/Nano characterizations were performed by atomic force microscopy (AFM, NT-MDT) (Figure S4), by a Scanning Electron Microscope (HIROX, SH-5000M) and a Transmission Electron Microscope (JEOL, 2100F) equipped with an energy-dispersive X-ray spectrometer operating at 200 kV.

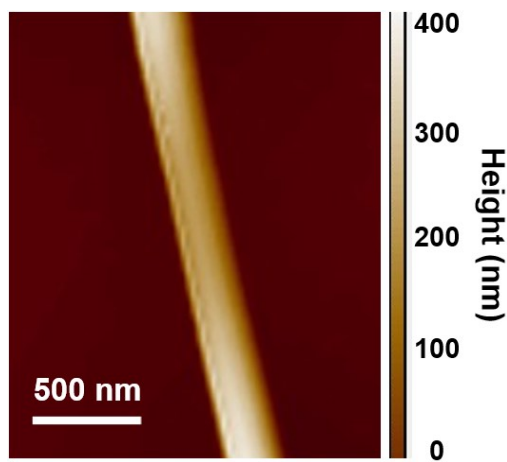

**Figure S4.** Atomic Force Microscope image of 300-nm-diameter nano-waveguide. The root-mean-square roughness of a 1 μm × 2 μm area is about 2 nm.

**Optical Characterization:**

Fluorescence micrographs were obtained using a microspectrometer system (CRAIC, 20/20 PV, USA) under the illumination excitation of a 420 nm laser beam. The orientation and position of tapered silica nanofiber with respect to hybrid nano-waveguide were carefully controlled by three-axis micromanipulator (Kohzu Precision, resolution of 50 nm) equipped with two tungsten probes (tip diameter of 200 nm) under optical microscopy. Micro-PL measurements were performed by using the microspectrometer system with spectral acquisition area of 1 × 1 μm$^2$. The spatial and temporal PL mapping were recorded by a confocal Raman microscope (HORIBA, XploRA) with a 532 nm laser excitation. The spot size of focused laser beam was about 300 nm.

**Wave-guiding properties:**

The light transport properties of the hybrid nano-waveguides can be described by modelling them as a cylindrical system with diameter, $D = 300$ nm. First, the fractional mode power of the fundamental mode, $\eta$, with wavelength, $\lambda$, can be calculated through the following relation [1]:

$$\eta = 1 - \frac{\left(2.405 \exp\left[-\frac{\lambda}{\pi D \sqrt{n_{hw}^2(\lambda)-1}}\right]\right)^2}{\left(\frac{\pi D}{\lambda}\sqrt{n_{hw}^2(\lambda)-1}\right)^3} \tag{S1}$$

Where $n_{hw}(\lambda)$ is the wavelength-dependent refractive index of the hybrid waveguide, here assumed to be mainly determined by the PVP matrix, which is given by: $n_{hw}(\lambda) = 1.5151 + 0.00279 \times \lambda^{-2} + 5.0756 \times 10^{-4} \times \lambda^{-4}$ [2]. Equation (S1) gives the fraction of the power of the fundamental guided mode that is confined within the fiber core, which is related to the fraction of fluorescence power emitted by the $MoS_2$ quantum dots (QD) that is guided through the PVP waveguide. Figure S5 shows the dependence of $\eta$ on the wavelength of the propagating light, evidencing that the confinement of more than 60% of the fundamental mode in the nano-waveguide core for wavelength < 650 nm.

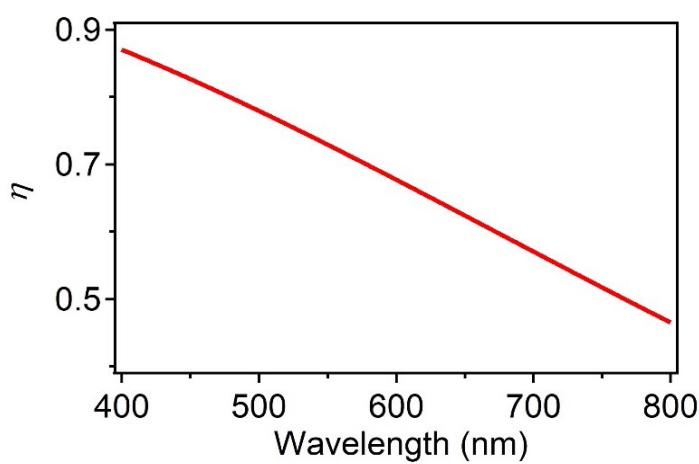

**Figure S5.** Calculated fractional mode power for the fundamental mode ($\eta$), as a function of the wavelength for propagating light in hybrid waveguide.

As a further analysis of the nano-waveguides, the propagation losses can be considered. Various effects can contribute to the measured propagation losses, $\alpha$, which can be summarized as:

$$\alpha = \alpha_{bulk} + \alpha_{Surf} + \alpha_{sub} \quad (S2)$$

Where $\alpha_{bulk}$, $\alpha_{Surf}$, and $\alpha_{sub}$ are the contribution to the propagation losses due to the bulk absorption and Rayleigh scattering ($\alpha_{bulk}$), to the scattering from surface defects ($\alpha_{Surf}$) and to coupling to the substrate ($\alpha_{sub}$), respectively. The losses due to the Rayleigh scattering from surface defects, $\alpha_{Surf}$, can be estimated by the expression derived for slab waveguides [3]:

$$\alpha_{Surf} = \left(\frac{4\pi}{\lambda}\right)^2 \left(\frac{\cos^3 \theta_p}{2\sin \theta_p}\right)\left(\frac{\sigma_R^2}{D+l_p}\right) \quad (S3)$$

where $\theta_p$ is the angle at which a mode can propagate with wavelength $\lambda$ that for PVP waveguides is $\theta_p > 41°$, $\sigma_R$ is associated with the fiber surface roughness, which was estimated to be 2 nm by atomic force microscopy (Figure S4), and $l_p$ is the propagation depth of a guided mode into the surrounding medium, that for the nano-waveguides here considered can be considered of the same order of $D$. $\alpha_{Surf}$ is of the order of 44 dB cm$^{-1}$ (0.0044 dB μm$^{-1}$), while $\alpha_{sub}$ is not expected to contribute significantly because of the use of a low refractive index substrate made of a layer of magnesium fluoride. Therefore, the main contribution to the propagation losses are due to the bulk of the hybrid nano-waveguides, namely the self-absorption and Rayleigh scattering from the embedded MoS$_2$ quantum dots.

**Numerical simulation:**

The numerical simulation for the reported geometry is presented in Figure S6. A nanofiber of 200-nm diameter is crossed with nano-waveguide of 300-nm diameter at 55-degree angle. The refractive index of the silica nanofiber and hybrid nano-waveguide are assumed to be 1.46 and 1.53, respectively, at the operation wavelength ($\lambda$) of 532 nm. In the simulation, the light sources of Gauss beam are selected to be polarized in the x-direction and y-direction, respectively, and sent into the nanofiber with its axis along +z direction. The hybrid nano-waveguide was supported by a low refractive index layer made of magnesium fluoride.

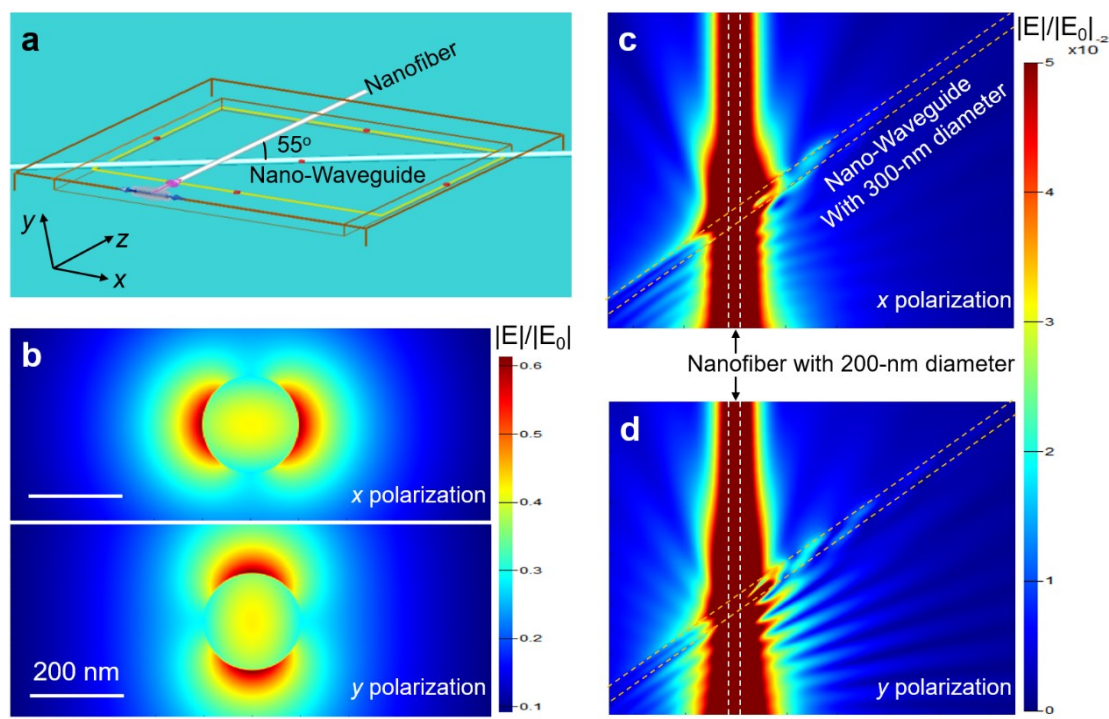

**Figure S6.** Numerical simulation for the reported geometry. (a) Structure model of a silica nanofiber crossed with a hybrid nano-waveguide. Gauss beam of 532-nm-wavelength is launched into the nanofiber from the left side. (b) Sources of $x$ polarization and $y$ polarization in the 200-nm-diameter nanofiber. The mesh accuracy is 2 nm. (c, d) Electrical field distribution of the reported geometry for $x$ polarization (c) and $y$ polarization (d), respectively. The mesh accuracy is 7 nm. The 532-nm-wavelength light is launched into the nanofiber from the bottom-to-top.

**White light generation:**

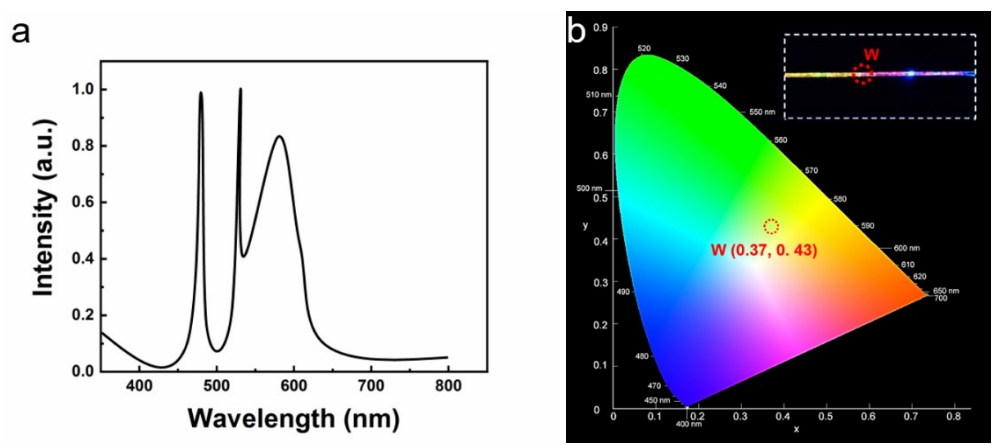

**Figure S7.** (a) Normalized PL spectrum and (b) chromatic coordinates (0.37, 0.43) of white light part, as indicated by red dashed circle labelled with "W" in the inset. Inset micrograph shows the position of white light generation in hybrid nano-waveguide, labelled with "W". The spectrum of white light part was measured by microspectrometer without using any filters on the path of the collection of the emission. Two sharp peaks at 473 and 532 nm is due to the excitation laser.

**References**

(1)   Johnson, J. C.; Yan, H.; Yang, P.; Saykally, R. J. Optical Cavity Effects in ZnO Nanowire Lasers and Waveguides. *J. Phys. Chem. B* **2003**, *107*, 8816-8828.

(2)   König, T. A. F.; Ledin, P. A.; Kerszulis, J.; Mahmoud; M. A.; El-Sayed, M. A.; Reynolds, J. R.; Tsukruk, V. V. Electrically Tunable Plasmonic Behavior of Nanocube-Polymer Nanomaterials Induced by a Redox-Active Electrochromic Polymer. *ACS Nano* **2014**, *8*, 6182-6192.

(3)   Tien, P. K. Light Waves in Thin Films and Integrated Optics. *Appl. Opt.* **1971**, *10*, 2395-2413.